\documentclass{aa501}
\usepackage{graphicx,natbib}
\bibpunct{(}{)}{,}{a}{}{,}
\begin{document}
\title{Radiative Hydrodynamic Modeling of the 
   Bastille-Day Flare (14 July, 2000): I. Numerical Simulations}
\author{D. Tsiklauri \inst{1} \and M.J. Aschwanden \inst{2}
\and V.M. Nakariakov \inst{3} 
\and T.D. Arber \inst{3} }
\offprints{David Tsiklauri, \\ \email{D.Tsiklauri@salford.ac.uk}}
\institute{Joule Physics Laboratory,
School of Computing, Science \& Engineering,
University of Salford, Salford, M5 4WT, England, UK
\and Lockheed Martin, Advanced Technology Center
Solar \& Astrophysics Laboratory, Dept. L9-41, Bldg.252
3251 Hanover Street, Palo Alto, CA 94304, USA 
 \and  Physics Department, University of Warwick, Coventry,
   CV4 7AL, England, UK }
\date{Received ???  2004 / Accepted ??? 2004}

\abstract{A 1-D loop radiative hydrodynamic model that incorporates
the effects of gravitational stratification,
heat conduction, radiative losses, external 
heat input, presence of helium, and Braginskii viscosity 
is used to simulate elementary flare loops. The physical parameters
for the input are taken from observations of the Bastille-Day flare of 2000 July 14.
The present analysis shows that: 
(a) The obtained maximum values of the electron density
can be considerably higher ($4.2\times 10^{11}$ cm$^{-3}$ or more) 
in the case of
footpoint heating than in the case of apex 
heating ($2.5\times 10^{11}$ cm$^{-3}$).
(b) The average cooling time after the flare peak takes less time in the case of 
footpoint heating than in the case of apex heating. 
(c) The peak apex temperatures are significantly lower
(by about 10 MK) for the case of footpoint heating than for apex heating
(for the same average loop temperature of about 30 MK).
This characteristic allows us to discriminate between different
heating positioning.
(d) In both cases (of apex and footpoint heating),
the maximum obtained apex temperature $T^{max}$  
is practically independent of the heating duration $\sigma_{t}$, 
but scales directly with the heating rate $E_{H0}$.
(e) The maximum obtained densities at the loop apex, 
$n_e^{max}$, increase with the heating rate $E_{H0}$ and heating duration
$\sigma_{t}$ for both footpoint and apex heating. 
In Paper II we will use the outputs of these hydrodynamic simulations, which
cover a wide range of the parameter space of heating rates and durations, 
as an input for forward-fitting of the multi-loop arcade of the 
Bastille-day flare.
 }
\titlerunning{Radiative HD Modeling of solar flares ...}
\authorrunning{Tsiklauri et al.}
\maketitle

\section{Introduction}

Solar flares are complex systems that involve many magnetic field lines and thus 
can rarely be represented by a single flux-tube. During the Bastille-Day 2000 July 14
flare for instance, which exhibits a classical double-ribbon flare configuration,
an ensemble of over 200 individual post-flare loops has been identified
(Aschwanden \& Alexander 2001). Each of these individual loops has its own 
hydrodynamic evolution during a flare, occurring in magnetic flux systems that
are thermodynamically
isolated from each other and have their own independent timing and physical
parameters. Hydrodynamic modeling of flare loops, however, have been performed
for single flare loops \citep{m87, m89}, but only few MHD simulation
studies have been orchestrated in a multi-loop configuration \citep{h97, h98}.
Even the multi-loop simulations have been designed only in the simplest
way, by assuming regular spacing and time intervals to merely mimic the
superposition effect, but no detailed fitting of the observed spatial configurations
and timing has ever been attempted. A rigorous hydrodynamic modeling effort for
complex large flares would be extremely valuable to constrain the total energy
budget, the heating functions, the fractal geometric structure, the plasma filling 
factors, and the spatio-temporal organization of {\it unsteady} \citep{p00} or
{\sl impulsive bursty magnetic reconnection} processes \citep{l82}, which are likely 
to occur in large double-ribbon flares because of the large shear and resulting 
tearing mode instability \citep{s66}. 

In this series of papers we present
a method of radiative hydrodynamic
modeling of large, complex, multi-loop flares. In this Paper I we perform numerical
simulations with a 1-D hydrodynamic code to obtain the temperature evolution 
$T_e^{max}(t), T_e^{avg}(t)$ and density evolution $n_e^{max}(t)$ in a large
parameter space of heating functions. We vary the maximum heating rate,
heating duration, and location (footpoint, apex) of the heating functions
$E_H(t)$. In Paper II we parameterize the results, suitable to forward-fitting of a
multi-loop system and fit the multi-wavelength data of the Bastille-day flare
2000 July 14 using TRACE, Yohkoh/SXT, HXT, and GOES data. In subsequent papers
we plan to extend the hydrodynamic results of this flare to constrain magnetic
modeling, magnetic reconnection geometries, and particle acceleration processes.

\section{The Radiative Hydrodynamic Model}

In order to describe plasma dynamics in a coronal loop,
we solve numerically 1D radiative hydrodynamic equations 
(infinite magnetic field approximation) 
that resemble closely the Naval Research Laboratory Solar Flux
Tube Model (SOLFTM) \citep{m87}. 
The numerical code that we use is a 1D version of
the Lagrangian Re-map code \citep{arber01} with radiative
loss limiters. 
As in SOLFTM the coronal loop is connected with a 
dense, cold (10000K), 5 Mm thick plasma region, that mimics
the chromosphere, and which because of its large
density provides sufficient amount of matter
to fill the loop during the flare.
The use term of 'chromosphere' should be taken with caution,
as we do not consider realistic one (with ionizations, 
proper radiative transfer, etc.).
Instead, we simply mimic chromosphere by considering
it as a dense, cold (10000K) plasma, which provides
source of matter inflow into the loop during the flare, while we
use radiative loss limiters to prevent plasma from
catastrophic cooling, as we use optically thin plasma
approximation ($\propto n_e^2$).
In such approach, specific initial conditions
in region connecting corona to the chromosphere and chromosphere itself
have little or no effect on the corona dynamics
as they are rapidly modified, self-consistently,
according to the radiative hydrodynamics equations.
In brief, our model includes: the effects of gravitational
stratification,
heat conduction, radiative losses, added external 
heat input, presence of helium, hydrodynamic 
non-linearity, and Braginskii bulk viscosity.
For the radiative loss function we use
the following parameterization,
$$
L_r(T)=n_e^2 
\left\{ \matrix{10^{-26.60} \; T^{1/2} & T>10^{7.6}  \cr
10^{-17.73} \; T^{-2/3} & 10^{6.3} < T < 10^{7.6}  \cr
10^{-21.94}  & 10^{5.8} <T < 10^{6.3}  \cr  
10^{-10.40} \; T^{-2} & 10^{5.4} <T < 10^{5.8}  \cr   
10^{-21.2}  & 10^{4.9} <T < 10^{5.4}  \cr      
10^{-31} \; T^{2} & 10^{4.6} < T < 10^{4.9}  \cr     
10^{-21.85}  &$ $10^{4.3} < T < 10^{4.6}  \cr     
10^{-48.31} \; T^{6.15} & 10^{3.9} < T < 10^{4.3}  \cr      
10^{-69.90}\; T^{11.7} & 10^{3.6}<T < 10^{3.9} \cr} \right.
$$
which is an extension of \citet{rtv78} compiled from other
sources \citep{p82,pe82}.

\section{Numerical results}

We start numerical simulations from the following
configuration: we take a semicircular loop with 
a length of $L=55$ Mm
(which corresponds to an average loop arcade radius of $r=17.5$ Mm,
as derived by \citet{aa01}, Table III, for the
Bastille-day flare). We keep the coronal part of the loop initially
at a temperature of 1 MK and at a mass density of
${\rho} = \mu m_p n_e = 6.6 \times 10^{-16}$ g cm$^{-3}$ (at the loop apex), 
for a helium-to-hydrogen number density ratio of 0.05,
i.e., with a mean molecular weight of 
$\mu = (1 + 0.05\times 4)/(1+0.05\times 2)=1.1$,
this corresponds to an electron density of 
$n_e=3.6\times 10^8$ cm$^{-3}$.
The resolution in all our numerical runs was fixed to
1000 grid points, which were distributed non-uniformly
in order to properly resolve strong gradients
in the region connecting corona to the chromosphere. As a convergence test,
runs with 3000 grid points were made, which showed no
difference to the case with 1000 grid points, thus confirming sufficient
numerical resolution in our simulations.

\subsection{Heating Function}

The heating function in flare loops is probably quite different
from that of non-flaring loops. Flare loops are filled by up-flowing
heated plasma from the chromosphere, once the chromospheric footpoints
become impulsively heated from precipitating non-thermal particles 
and/or downward propagating hot thermal conduction fronts from the
coronal reconnection site. Chromospheric evaporation seems to be
the main matter inflow 
source for flare loops. The heating function has
therefore to accommodate very localized heating at 
the footpoints \citep{tn01,asa01}.
On the other side, reconnection outflows from the reconnection
region contain also heated plasma and heat the flare loops from
the apex side, for instance in the standard reconnection model 
of \citet{kp76}, which seems to fit the magnetic configuration 
of the Bastille-day flare to first order \citep{aa01}.
Thus, to allow for both options, we have performed numerical 
simulations for both cases separately, i.e., for apex and 
footpoint heating functions.

We have used the following heating function in our simulations:
\begin{equation}
   	E_H(s,t)=E_H^s(s) E_H^t(t) =
	E_0 \biggl[ \exp\left(-{{(s+s_0)^2}\over{2 \sigma_s^2}}\right)+ 
\end{equation}
$$	
	\exp\left(-{{(s-s_0)^2}\over{2 \sigma_s^2}}\right) \biggr] 
	\times \left[1+\alpha Q_p  
	\exp\left(-{{(t-t_p)^2}\over{2 \sigma_t^2}}\right) \right]
$$	

Here, $E_H^s(s)$ and $E_H^t(t)$ are the spatial and temporal
parts of the heating function, which are
taken to be independent of each other for simplicity.
$E_0$ is the heating rate in units of [erg cm$^{-3}$ s$^{-1}$].
The positions $s = \pm s_0$ are the locations with the maximum
heat deposition, i.e., $s_0=0$ for apex heating. The
heat deposition length scale is called ${\sigma}_s$ (i.e., the spatial 
width of the Gaussian).
The temporal part of the heating function is similar to one
used by \citet{aa01} (cf. their Eq.~(31)), where $t_p$ is the flare
peak time and $\sigma_t$ is the  duration of the flare (i.e., the temporal 
 width of the Gaussian). However, our choice  of the temporal
part of the heating function is such that there is a small background 
heating present at all times (either at footpoints or
the loop apex) which ensures that in the absence of flare heating
(when $\alpha=0$, a parameter that determines the flare heating amplitude),
the average loop temperature stays at 1 MK.
For easy comparison between apex and footpoint heating
cases we fix the flare heating amplitude $Q_p$ at a given different 
value in each case. This ensures that the average loop temperature
peaks at about at the observed value of 30 MK in {\it both cases}, 
when the the flare heating is on ($\alpha=1$).
Then we vary also $\alpha=0.01, 0.05, 0.25, 0.5, 1.0$ to obtain different
apex temperatures $T^{max}$.
Hereafter, we denote the (flare) heating rate by $E_{H0}=E_0 \alpha Q_p$.

In all our numerical runs presented here,  $1/(2 \sigma_s^2)$
was fixed to a value of 0.01 Mm$^{-2}$, which gives a heat deposition length scale 
of $\sigma_s=7$ Mm. This is a typical value determined from
observations \citep{abk02}. The flare peak time was fixed in our numerical
simulations at $t_p=822$ s. The time step of data visualization
(which in fact is much larger that the actual time step, 0.035 s, in
the numerical code) was chosen to be $\Delta t=10$ s.

\subsection{Case of Apex Heating}

In the case of apex heating we fix $s_0=0$ Mm in Eq.(1).
Initially we run our numerical code without flare heating,
i.e. we put $\alpha=0$ (in this manner we turn off flare 
heating). 
In all our numerical runs we keep background heating
on at all times, so that flare occurs starting
from a steady, equilibrium loop.
The result of this simulation is presented in
Fig.~1. 
\begin{figure*}
\centering
 \includegraphics[width=17cm]{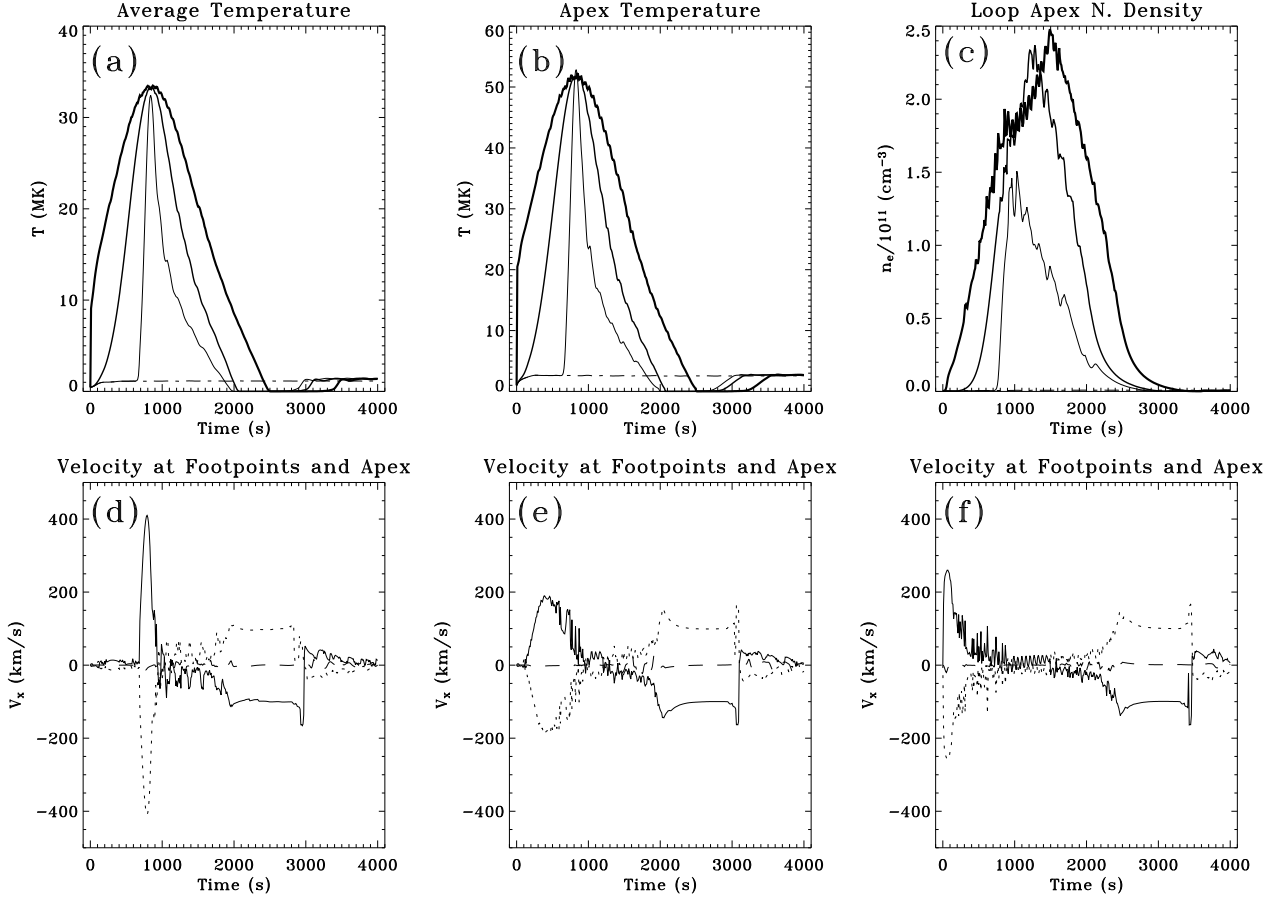}
\caption{{\bf (a)} Average temperature as a function of time in
the case of apex heating. 
Dash-dotted curve corresponds to the case of background
heating ($\alpha=0$). The thin, thick and thickest solid
lines correspond to the cases of background plus
flare heating ($\alpha=1$) for different flare durations, 
$\sigma_t=41, 164, 329$ s, respectively.
{\bf (b)} The same as in {\bf (a)}, but for the temperature
at the apex. {\bf (c)} The same as in {\bf (a)}, but for the
number density at the apex.
{\bf (d)} Velocities versus time at footpoints ($s=\pm 27.5$),
solid and dotted curves, 
and apex ($s=0$), dashed curve, for the case of  
background plus
flare heating ($\alpha=1$) for the flare duration, 
$\sigma_t=41$ s.
{\bf (e)} The same as in {\bf (d)}, but for $\sigma_t=164$ s.
{\bf (f)} The same as in {\bf (d)}, but for $\sigma_t=329$ s.
\label{fig1}}
\end{figure*}
Dash-dotted curves in panels (a) and (b)
show the evolution of the average and apex temperatures in time.
Since the observational data of the 
Bastille day flare, namely the temperature
evolution (cf. top left panel of Fig.~11 in \citet{aa01}),
was derived for the entire field of view of the
instrument, we found that it is useful, in addition 
to the apex temperature, to track also 
the {\it average} loop temperature, defined
as the sum of temperatures along the loop divided by
the number of grid points at a given snapshot. 
Since the outputs from our hydrodynamic simulations
will be used to model the observational data of the 
Bastille-day flare in paper II,
$E_0$ was fixed to 0.002 erg cm$^{-3}$ s$^{-1}$ 
in all runs of this subsection, which insures 
that the average loop temperature stays at 1 MK 
in the absence of flare heating (cf. Fig.~1a).
Note that the apex temperature in this case tends
to have a higher asymptotic value of 2.5 MK (cf. Fig.~1b).
The obtained asymptotic value of the loop apex number
density is $n_e = 0.9\times 10^{9}$ cm$^{-3}$ in this case
(and thus is not visible in Fig.~1c).

Then, we run our numerical code with flare heating,
i.e. we put $\alpha=1$ and fix $Q_p=3 \times 10^4$,
so that it yields peak average temperature of about 30 MK, as observed
during the Bastille-day flare \citep{aa01}, and then we vary the 
duration of the heating phase to $\sigma_t=41$, 164, 329 s (i.e., the
temporal width of the Gaussian).
Note that it is the combined action of the heating rate $Q_p$, 
the flare heating amplitude $\alpha$,
and the duration $\sigma_t$ of the flare, that determine the
effect of flare heating, i.e. the term
$\alpha Q_p  \exp\left(-{{(t-t_p)^2} / {2 \sigma_t^2}}\right)$
in Eq.(1), compared with the background heating rate ($\alpha =0$). 
The results are presented in Fig.~1. 
There are several noteworthy features in this graph:
(1) As expected, an increase of the heating time interval 
$\sigma_t$ yields an increase 
of the flare duration (cf. Fig.~1(a,b)).
(2) However, an increase in the duration of the heating phase
does not affect the obtained maximum values of the
average nor the apex temperature (cf. Fig.~1a and 1b).
This is counter-intuitive, because we expect that 
the amount of deposited heat increases with the heating duration.
This invariance of the obtained flare temperatures with
respect to duration of the flare could probably be explained by
some form of saturation (or balance) in the combined 
action of losses in the system -- heat conduction and
radiative losses -- and flare heat input.
(3) An increase in the duration of heating should naturally
result in an increase of the plasma density in the loop.
For the long duration flares (the thickest solid
line in Fig.~1c) there are three clearly different physical
regimes which yet have to be properly identified.
(4) Very useful information can be extracted from the 
velocity outputs at the footpoints ($s=\pm 27.5$) and at the apex ($s=0$),
as function of time (cf. Fig.~1d, 1e, 1f).
We find that strong up-flows (up to 400 km s$^{-1}$) 
are present at the footpoints of the flare loops during the flare onset, 
as the deposited heat (delivered by conduction from the apex to the 
footpoints) causes material evaporation from
the dense chromosphere. The late up-flow phase is followed
by an oscillatory phase with typical
amplitudes of a few tens of km s$^{-1}$, which in turn is followed
by down-flows (up to 100 km s$^{-1}$), when plasma
is drained out of the loop
(5) Note that the velocity dynamics at opposite footpoints
remains perfectly symmetric at all times (cf. Fig.~1d, 1e, 1f), 
while net flow through the apex remains zero at all times
as expected. This is due to the symmetry of the problem.
(6) The peak up-flow velocity during the onset of a flare
decreases with the increase of the duration of the flare
(for $\sigma_t=41$s). In Fig.~1d the peak up-flow velocity
is about 400 km s$^{-1}$, while for $\sigma_t=164$s (Fig.~1e),
it is about 200 km s$^{-1}$). 
This may seem counter-intuitive at first glance,
since we would expect that the duration of the
up-flow itself increases with increasing heating duration
$\sigma_t$. Thus, the net material evaporated into the
loop from chromosphere still increases with the
increase of the heating duration, so that the
densities obtained at the apex increase (cf. Fig.~1c).
(7) Yet another interesting observation is that
the time instances when down-flows abruptly
end after the flare (cf. Fig.~1d, 1e, 1f), these times 
correspond exactly to the same time instances when
both the average and apex temperature curves reach 
these asymptotic values (thin, thick, and very thick
solid lines that join the dash-dotted lines in Figs. 1a and 1b),
signaling the end of the flare heating phase and the onset of a steady state.
In fact, statements (6) and (7) are true for all numerical 
runs performed (see below).

In a next step we investigate the effect of the
flare heating amplitude, by fixing the duration of the flare
$\sigma_t=41$ s and varying $\alpha=0.01,0.25,1.0$
(again with fixed $Q_p = 3 \times 10^4$).
The results of these numerical runs are presented in Fig.~2.
\begin{figure*}
\centering
 \includegraphics[width=17cm]{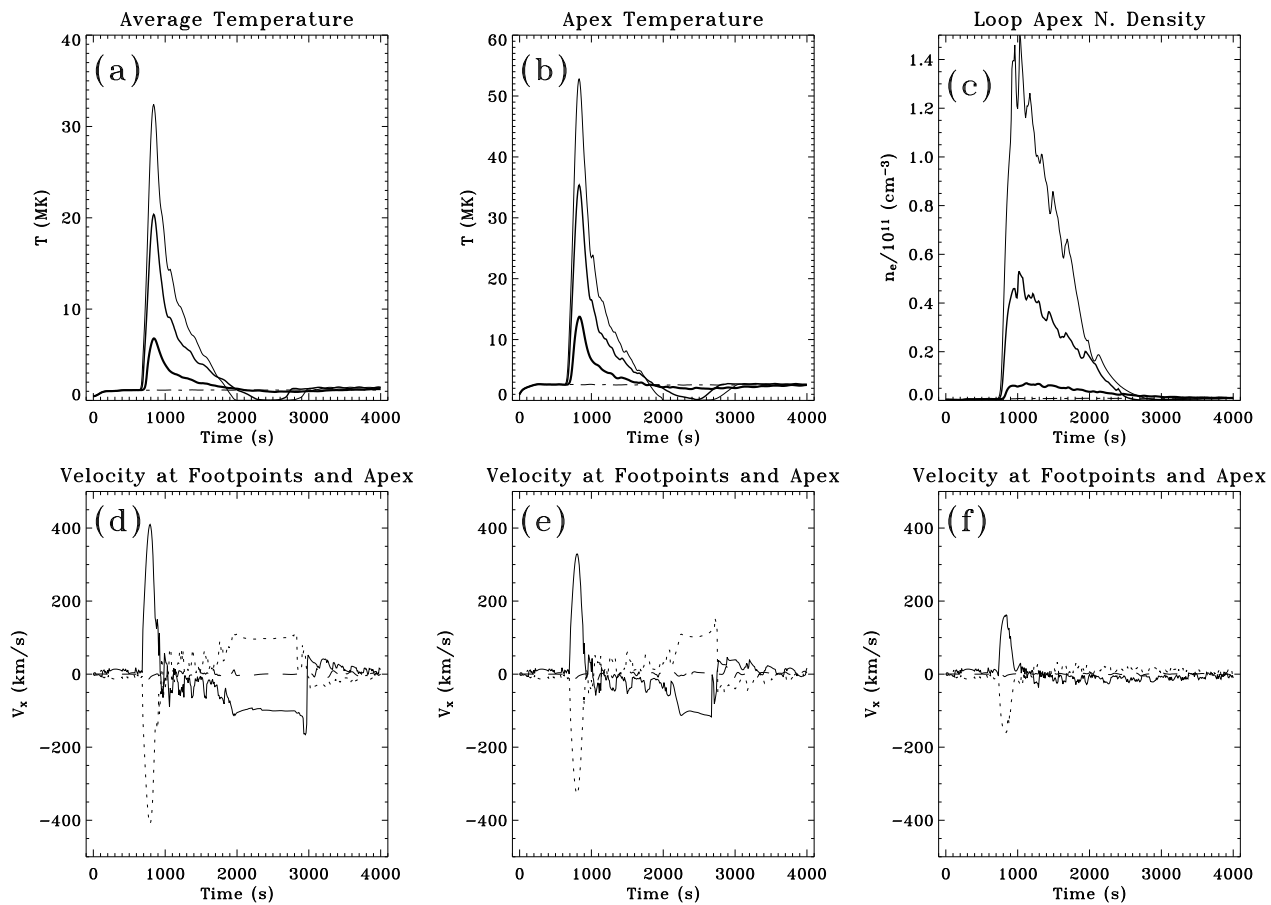}
\caption{{\bf (a)} Average temperature as a function of time in
the case of apex heating. 
Dash-dotted curve corresponds to the case of background
heating ($\alpha=0$). The thin, thick and thickest solid
lines correspond to the cases of background plus
flare heating ($\alpha=1$) for different flare peak amplitudes, 
$\alpha=0.01,0.25,1.0$, respectively (these correspond to 
flare heating rates $E_{H0}=0.6, 15.0, 60.0$
erg cm$^{-3}$ s$^{-1}$).
{\bf (b)} The same as in {\bf (a)}, but for the temperature
at the apex. {\bf (c)} The same as in {\bf (a)}, but for the
number density at the apex.
{\bf (d)} Velocities versus time at footpoints ($s=\pm 27.5$),
solid and dotted curves, 
and apex ($s=0$), dashed curve, for the case of  
background plus flare heating (flare duration $\sigma_t=41$ s) 
for the flare peak amplitude, $\alpha=1.0$ (which corresponds to 
flare heating rate $E_{H0}=60.0$ erg cm$^{-3}$ s$^{-1}$).
{\bf (e)} The same as in {\bf (d)}, but for $\alpha=0.25$ 
($E_{H0}=15.0$ erg cm$^{-3}$ s$^{-1}$).
{\bf (f)} The same as in {\bf (d)}, but for $\alpha=0.01$
($E_{H0}=0.6$ erg cm$^{-3}$ s$^{-1}$). \label{fig2}}
\end{figure*}
There are several interesting features in this graph:
(1) As expected, a decrease of $\alpha$ results in a decrease of
the obtained flare temperature (cf. Fig.~2a and 2b).
(2) Also, a decrease of the flare heating amplitude
does not affect the duration of the flare (cf. Fig.~2a and 2b).
(3) In addition, a decrease of the flare heating amplitude 
naturally results in a decrease of the plasma density 
in the loop. This is understandable as the less deposited
heating rate causes less material evaporation form the chromosphere
and, hence, produces less dense and cooler loops during the flare (cf. Fig.~2c, 2a, and 2b).
(4) Again, as in Fig.~1d, 1e, 1f, we studied the resulting  
velocities at the footpoints ($s=\pm 27.5$ Mm) and apex ($s=0$)
as function of time (cf. Fig.~2d, 2e, 2f).
Our simulations show that the 
peak up-flow velocities during the onset of the flare
decrease with decreasing heating amplitudes.
For $\alpha=1.0$ (Fig.~1d), the peak up-flow velocity
is about 400 km s$^{-1}$, while for $\alpha=0.01$ (Fig.~1f),
it is about 160 km s$^{-1}$). 
Also, note that the 4-fold decrease in the
heating flare amplitude (Fig.~1e) still results in
down-flows of the order of 100 km s$^{-1}$, which
are typical for most runs, while further decrease in $\alpha$
(Fig.~1d) causes an absence of any noticeable down-flows.

\subsection{Case of Footpoint Heating}

In the case of footpoint heating we fix $s_0=\pm 30$ Mm in Eq.(1), 
i.e. the (spatial) peaks of the heating are chosen to be at the bottom 
of the region connecting corona to the chromosphere (i.e. top of chromosphere).
As for the case of apex heating, 
we run initially our numeric code without flare heating
($\alpha=0$). The output from  this numerical run of the
code is presented in
Fig.~3. 
\begin{figure*}
\centering
 \includegraphics[width=17cm]{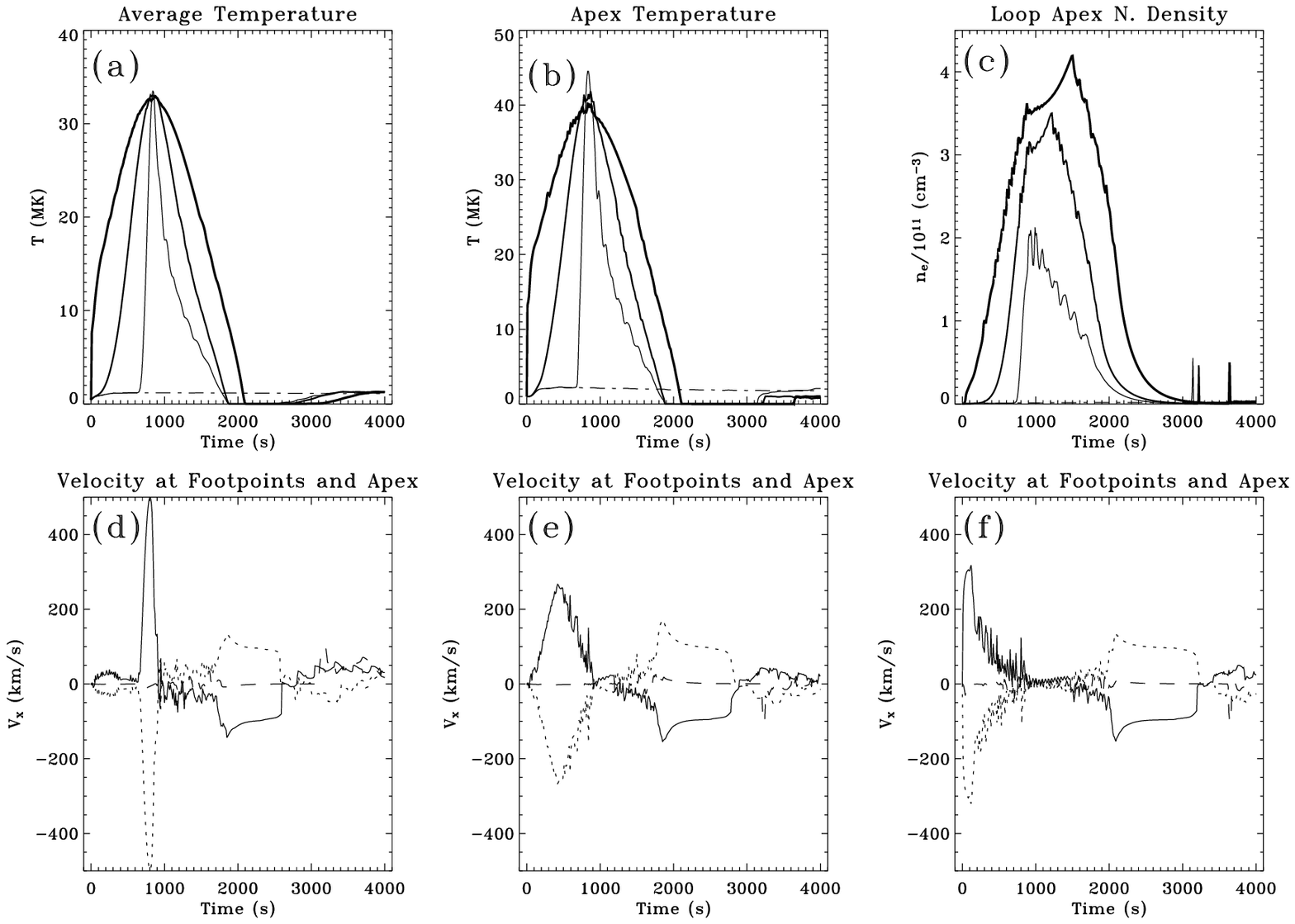}
\caption{The same as in Fig.~1, but for footpoint heating. 
 \label{fig3}}
\end{figure*}
$E_0$ is in all runs in this subsection fixed to a value of
0.01 erg cm$^{-3}$ s$^{-1}$, which insures that the average loop 
temperature stays at 1 MK in the absence of flare heating (cf. Fig.~3a).
Note that in the case footpoint heating, the amount of 
deposited heat required to keep the loop at an average temperature
of 1 MK is 5 times greater than in the case of apex heating
(recall that in the later case $E_0=0.002$ erg cm$^{-3}$ s$^{-1}$).
This increased heating is required in order
for the upward heat conduction to be sufficient enough to keep the loop
at this temperature.
The apex temperature in the case of footpoint heating approaches
an asymptotic value of 1.7 MK (cf. Fig.~3a), while
the obtained asymptotic value of the loop apex number
density is $0.02\times 10^{11}$ cm$^{-3}$ in this case
(not visible in Fig.~3c).
Note that the asymptotic value of the loop apex temperature, $T^{max}=1.7$ MK, 
is lower in the case of footpoint heating (due to the flatness of the temperature
spatial profile along the loop) compared with the case of apex heating ($T^{max}=2.5$ MK).
The larger amount of deposited heat in the case of
footpoint heating results in an asymptotic value
of the the loop apex density ($n_e=0.02\times 10^{11}$ cm$^{-3}$)
that is more than twice the value for 
apex heating ($n_e=0.009\times 10^{11}$ cm$^{-3}$).
This is due to the fact that the more deposited heat causes more
material evaporation from the chromosphere into the loop.

Yet another interesting point that only appears in the
case of footpoint heating is that there is a minimal
heat deposition length scale, $\sigma_s$, which allows
the existence of a steady loop at an average temperature
of 1 MK. The results presented in this subsection are
for $\sigma_s=7$ Mm. Smaller values 
cause a gradual formation of a condensation at the apex of the
loop (clearly seen in the animations, not included here), 
which then forms a prominence due to the thermal instability 
and causes an eventual disappearance of the loop.
This occurs even with a substantial increase 
of the heating amplitude $E_0$.
This is caused by the inefficiency of upward
heat conduction to keep the loop at a typical
coronal temperature of the order of $T\approx 1$ MK.

Our next step is to switch on flare heating as described in
previous subsection. The physical parameters used here are
also the same (apart from differences in the 
steady state as described above) for easy comparison.
Namely, we put $\alpha=1$, and fix $Q_p=1.5 \times 10^4$,
so that it yields an average peak temperature of about $T^{max}=30$ MK as
in the Bastille-day flare \citep{aa01}, and we vary duration of the flare
by changing the heating length scale to $\sigma_t=41, 164, 329$ s.
The results of this simulation are presented in Fig.~3.
Since, the dynamics of flare with variation of its
duration has been described in considerable detail
for the case of apex heating (Fig.~1), we focus now our
attention on differences between the two cases.
These can be summarized as follows:
(1) We observe (Fig.~3c) that 
the obtained maximum values of the density
are considerably higher ($n_e=4.2\times 10^{11}$ cm$^{-3}$) 
in the case of footpoint heating 
than in the case of apex heating ($n_e=2.5\times 10^{11}$ cm$^{-3}$),
(compare Fig.3c with Fig.1c).
This is due to the fact that footpoint heating is more
efficient in evaporating material from the region connecting corona 
to the chromosphere
and chromosphere itself, yielding denser loops during the flare.
In fact, previous hydrodynamic simulations 
assert that there is a maximum density limit, no matter how much 
heat is deposited. 
This limiting density is about $2\times 10^{11}$ cm$^{-3}$
(\citep{m89}, cf. their Fig.~2, after 60 s). 
\citet{fh90} get similar values.
An analogous problem exists in quite Sun loops. 
\citet{asa01} points out that the observed TRACE loops 
show a higher
density and pressure than expected from the RTV law, 
which we might call
the \lq\lq{over}-density loop problem".
However, as we can see in Fig.~3c, it is possible to break that
\lq\lq{density} limit" and the reason for this is that we have used
{\it footpoint} heating. 
In the case of footpoint heating it is possible to
evaporate more material from the chromosphere and
the region connecting corona to the chromosphere, 
where the density is orders of magnitude
higher than in the corona. 
In the case of apex
heating, in contrast, insufficient downward heat conduction
prevents significant evaporation.
Works of \citet{m89} and \citet{fh90} presumably were motivated
to use apex heating because of the standard reconnection model 
\citep{kp76}, which implies that the
heating function is localized in the loop apex
(actually, \citet{m89} used a
model in which the (collisional) heating by a beam of nonthermal electrons
injected at the apex followed the density evolution of the flare loop and so
varied in time in response to the evolution of the flare).
Consequently they obtain lower densities in the loop.
Similar results that by modeling a loop as set of
small scale, impulsively heated filaments one can generally
reproduce the observed flaring loops were obtained by
\citet{wwm03}.
(2) In the case of footpoint heating, a flare takes less time 
on average (compare Fig.3a and 3b with Fig.1a and 1b) --
so the cooling of the loop happens faster.
This can be explained by the following reason: 
At the initial stages of cooling
conductive losses dominate over radiative losses (when T $\geq$ 20 MK),
while the situation reverses (when T $\leq$ 20MK) as the loop cools
down and becomes less dense -- plasma is flowing out
of the loop (strong down flows were seen to accompany 
the cooling process in all of our
numerical simulations and this is also confirmed in a 
number of observations).
The time scale of conduction loss is proportional to the density,
while the time scale of radiative loss is proportional to the
inverse density  \citep{aa01}. Therefore, since the radiative losses
dominate over the heat conduction losses for most of the time,
it is natural to conclude that {\it the denser loops would cool
faster}. 
(3) In the case of footpoint heating the
peak apex temperatures ({\it corresponding to the same,
as in the case of apex heating, average temperature of 
about 30 MK}) are significantly lower (less by about 10 MK).
The observational data of the 
Bastille-day flare, namely the temperature
evolution (cf. top left panel of Fig.~11 in \citet{aa01}),
was derived for the entire field of view of the
instrument, i.e. it tracks the dynamics of the {\it average}
temperature. However, if one would have additional temperature dynamics 
at a loop given point such as at the apex,
{\it our simulations would allow to discriminate} 
between the different heating functions of the loop during
the flare.
(4) In the case of footpoint heating up-flow velocities 
are somewhat higher (compare Fig.3d, 3e, 3f with Fig.1d, 1e, 1f).

To complete the comparison between apex and
footpoint heating cases,  we investigate the effect of the 
flare heating amplitude, by fixing the duration of the flare to
$\sigma_t=41$ s, and vary $\alpha=0.01,0.25,1.0$
(again with fixed $Q_p=1.5 \times 10^4$).
The results of these numerical runs are presented in
Fig.~4 (see for comparison Fig.~2).
\begin{figure*}
\centering
 \includegraphics[width=17cm]{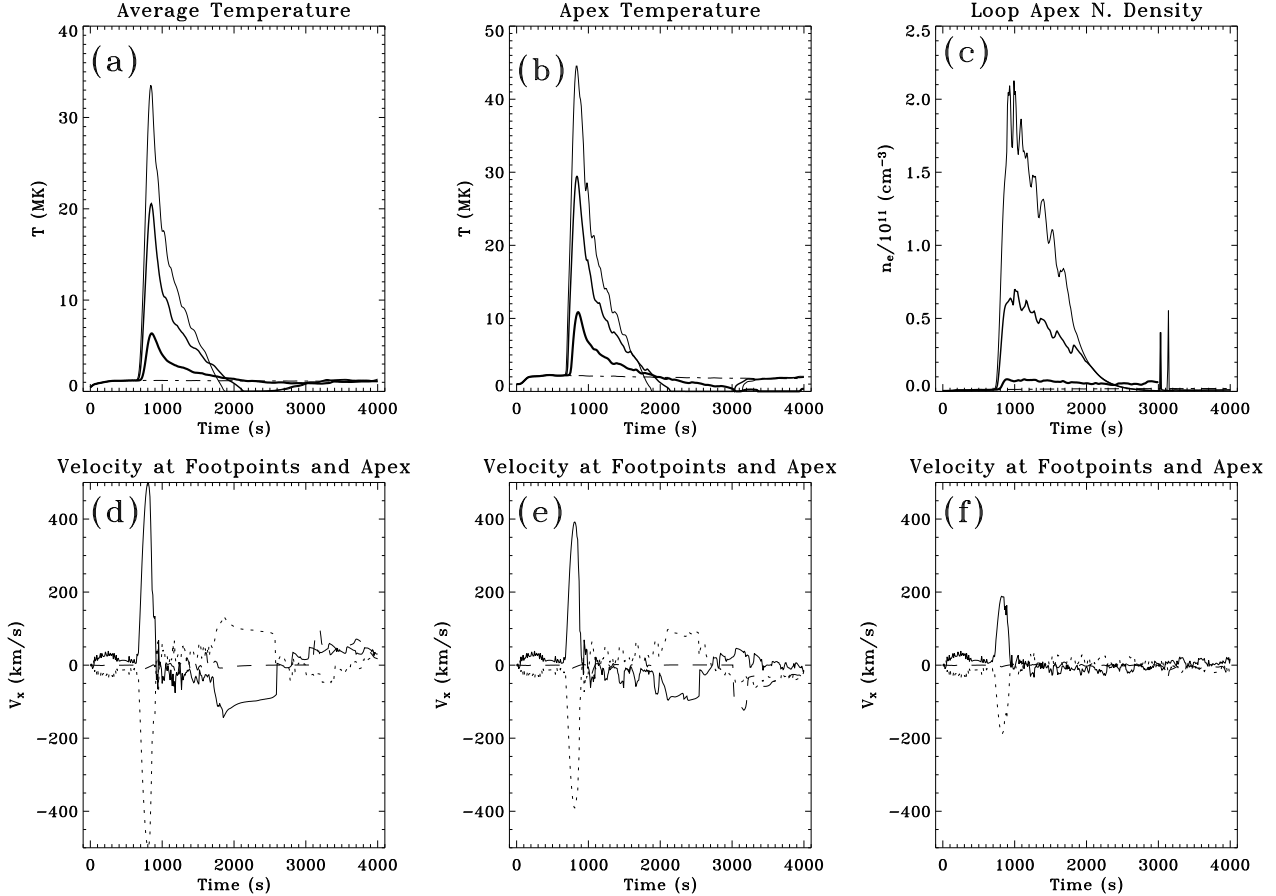}
\caption{The same as in Fig.~2, but for footpoint heating. 
However, the flare peak amplitudes $\alpha=0.01,0.25,1.0$ correspond 
here to the flare heating rates $E_{H0}=1.5, 37.5, 150.0$
erg cm$^{-3}$ s$^{-1}$).
\label{fig4}}
\end{figure*}
Most of the features seen in the case of apex heating
are still valid in this case. A notable difference is that, 
{\it in the considered parameter space},
spikes in the apex density time profiles occur (Fig.~4c),
which can be also seen in Fig.~3c after the flare.
In fact, these spikes are present in all runs with 
footpoint heating. They occur when dense blobs of
plasma (prominences), formed by the thermal instability,
swipe through the apex. One should mention, however, 
that a 1D code does not provide a fully adequate description
of prominences, since it ignores the finite magnetic
field tension (we use an infinite magnetic field limit in our code).
In fact, this is the reason why they do not stay 
steady at the apex, as the magnetic field in our case
cannot bend to provide cavity for a stationary prominence.
Yet another difference between the two cases are
the higher obtained densities and up-flow velocities
during the flare in the case footpoint heating
(due to more efficient evaporation) and lower obtained
apex temperatures (due to the spatial flatness of the heating
function along the loop).

\subsection{Parametric Study}

In this subsection we present a parametric study of the
problem by investigating the maximum obtained temperatures
and densities at the loop apex as a function 
of flare duration, $\sigma_t$ and the heating rate $E_{H0}=E_0 \alpha Q_p$.
We have performed simulations for $\sigma_t=41, 82, 164, 329$ s
and $\alpha=0.01, 0.05, 0.25, 0.5,  1.0$
with $Q_p=3 \times 10^{4}$ for the case of apex heating
and $Q_p=1.5 \times 10^{4}$ in the case of footpoint heating.
With appropriate multiplication by $E_0$ in each case
($E_0=0.002$ erg cm$^{-3}$ s$^{-1}$ for apex heating and 
$E_0=0.01$ erg cm$^{-3}$ s$^{-1}$ for footpoint heating),
the flare heating rates, $E_{H0}$, are,
$0.6, 3.0, 15.0, 30.0, 60.0$ erg cm$^{-3}$ s$^{-1}$ in the case
of apex heating and 
$1.5, 7.5, 37.5, 75.0, 150.0$ erg cm$^{-3}$ s$^{-1}$ in the
case of footpoint heating.
The results of these numerical runs are 
presented in Fig.~5. 
\begin{figure*}
\centering
 \includegraphics[width=17cm]{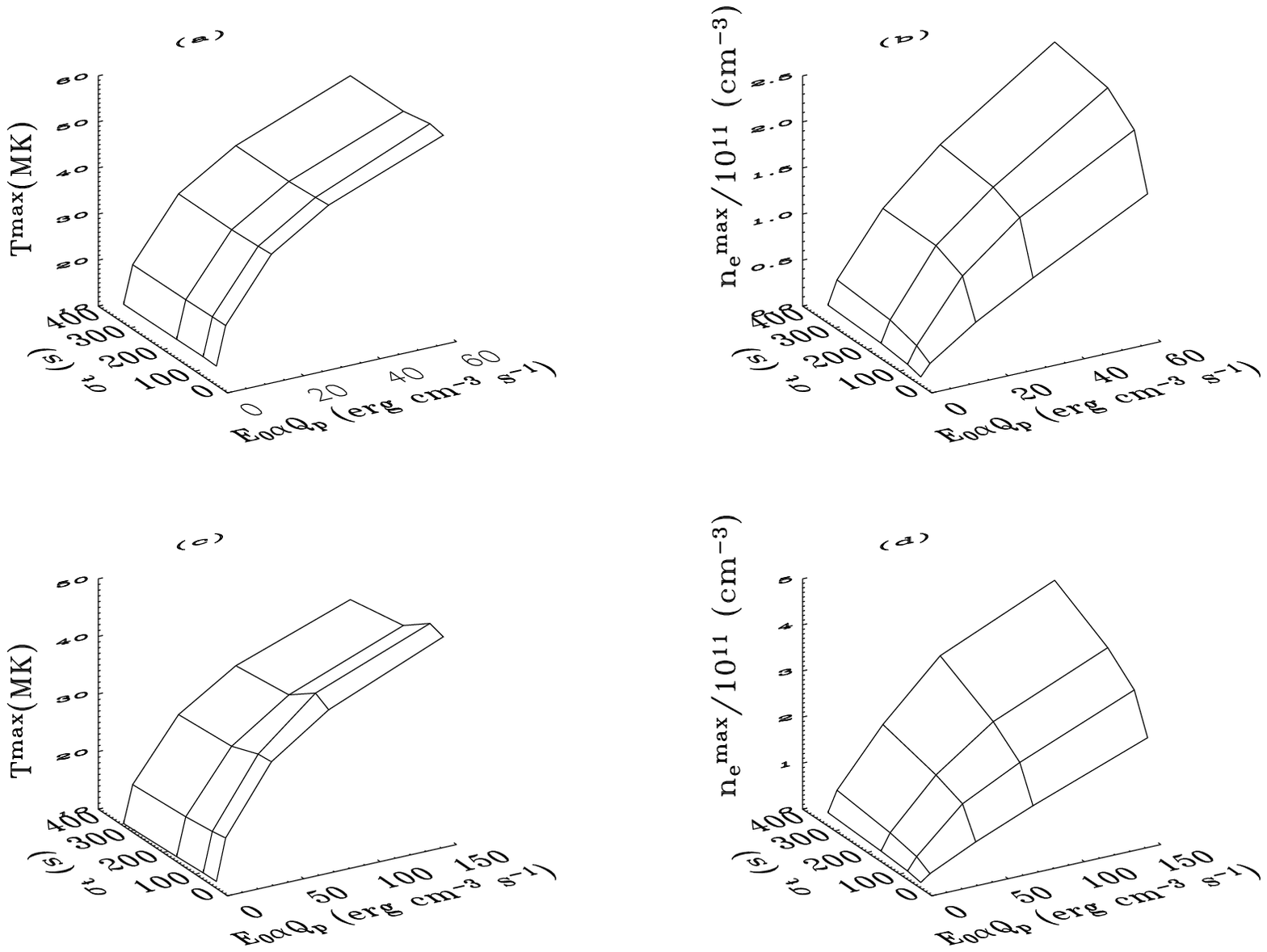}
\caption{ 
{\bf (a)} The dependence of 
maximum obtained temperatures at the loop apex as a function 
of flare duration, $\sigma_t$ and heating rate $E_{H0}$
for the case of apex heating.
{\bf (b)} The same as in {\bf (a)}, but for the maximum obtained
density in the case of apex heating.
{\bf (c)} The same as in {\bf (a)}, but for the case of footpoint
heating.
{\bf (d)} the same as in {\bf (b)}, but for footpoint heating.
\label{fig5}}
\end{figure*}
The actual data that was used to produce
Fig.~5 is given in Table 1.
In Figs.~5a and 5c we plot the maximum obtained 
temperatures, $T^{max}$ (MK),
at the loop apex in the case of apex and footpoint heating,
respectively. We gather from these plots that:
(1) In both cases $T^{max}$ is practically independent of
the flare duration $\sigma_t$ and it increases with the
increase of the flare heating rate $E_{H0}$.
(2) $T^{max}$ is up to 10 MK higher in the case of apex heating 
than in the case of footpoint heating.
This due to the steepness of the temperature
profile along the loop in the case of apex
heating (corresponding to the same average temperature
as in the case of footpoint heating).
In Figs.~5b and 5d we plot the maximum obtained densities, 
$n_e^{max}/10^{11}$ (cm$^{-3}$),
at the loop apex in the case of apex and footpoint heating,
respectively. It can be seen that
(1) In both cases $n_e^{max}$ increases with the
increase of both the flare heating rate $E_{H0}$ and
flare duration $\sigma_t$, and
(2) $n_e^{max}$ is about twice 
as high in the case of footpoint heating than in the case of apex
heating. This is due to more efficient material
evaporation from the chromosphere and 
the region connecting corona to the chromosphere 
in the case of footpoint heating.
We have also performed parametric study varying the loop
length. The results are presented in Table 2.
Note that for consistent comparison with other numerical runs,
in the case of footpoint heating, when varying
loop length, we have
shifted $s_0$ so that spatial maximum of the
heating function always stayed at the top 
of chromosphere.

\section{Conclusions}

In summary, we have used a  radiative
hydrodynamic numerical code
to simulate flares. The physical parameters of the input
were obtained from observations of the the Bastille-day flare \citep{aa01}.
Our simulations confirm the general picture of flare dynamics:
Transient heat deposition either at loop footpoints (with maximum heat 
input at the bottom of the region connecting corona to the chromosphere
or i.e. top of chromosphere)
or at the apex leads to an average loop 
temperature of $T_{avg} \approx 30$ MK first.
Then, evaporation of material from the chromosphere and 
the region connecting corona to the chromosphere
into corona ensues with up-flows in the order of 
a few hundreds of km s$^{-1}$. During the peak of
the flare, the combined action of heat input and conductive and
radiate loss yields an oscillatory flow pattern with
typical amplitudes of up to few tens of km s$^{-1}$.
Finally we enter a cooling phase, when down-flows in the
order of hundred km s$^{-1}$ can be seen as the
plasma drains out of the loop, ultimately reaching an 
equilibrium.

We have established the following:
\begin{enumerate}
\item In the case of
footpoint heating, the obtained maximum values of the density
are considerably higher ($4.2\times 10^{11}$ cm$^{-3}$ or more) than 
in the case of apex heating ($2.5\times 10^{11}$ cm$^{-3}$).
This is due to the fact that footpoint heating is more
efficient in evaporating material from the region connecting corona 
to the chromosphere
and chromosphere itself, which yields denser loops during the flare.
In the case of apex
heating, which was used to model flares, insufficient downward heating conduction
prevents significant material evaporation.
\item In the case of footpoint heating, as compared to the apex heating, 
on average cooling after the flare takes less time.
This due to the fact that 
the time scale of conduction loss is proportional to the density,
while the time scale of radiative loss is reciprocal to the
density  \citep{aa01}. Therefore, since radiative losses
dominate over heat conduction losses for most of the time,
it is clear that  the denser loops cool faster. 
\item In principle, {\it our 
simulations would allow to discriminate} between different heating 
functions of the loop during the flare, 
if  one would have temperature dynamics in a given point of the loop,
such as at the apex. This is based on our observation that
in the case of footpoint heating the
peak apex temperatures ({\it corresponding to the same,
as in the case of apex heating, average temperature of 
about 30 MK}) are significantly lower (less by about 10 MK).
\item In the case of footpoint heating, up-flow velocities 
are higher (roughly up to  100 km s$^{-1}$) than in the case of apex
heating due to more efficient evaporation.
\item In both cases (of apex and footpoint heating)
the maximum obtained temperature $T^{max}$ at the loop apex 
is practically independent of the 
heating duration $\sigma_t$, but it increases with 
higher heating rates $E_{H0}$.
\item  The maximum obtained densities at the loop apex
increase with the increase of both the flare heating rate $E_{H0}$ and
the heating duration $\sigma_t$, in the case of apex as well as
footpoint heating.
\item Varying the loop length (see Table 2) in the range of 
$L=(0.25,...,2.0)\times L_0$ (with $L_0=55$ Mm), we find 
(1) that the mean loop temperature averaged over the loop
length does not change dramatically, 
(2) that the loop apex temperature increases notably 
for longer loops only
for the case of apex heating, but much less for footpoint
heating, and (3) that the mean electron density decreases somewhat
with longer loops, i.e., $n_e/10^{11}$ (cm$^{-3}$)
$=10.45 L^{-0.362}$ for apex heating
and $n_e/10^{11}$ (cm$^{-3}$)
$=9.62 L^{-0.207}$ for footpoint heating.
Here, $L$ is loop length in Mm.

\end{enumerate}

In practically all of our numerical runs we have detected
quasi-periodic oscillations in all physical quantities.
In fact, such oscillations are frequently seen 
during solar flares (e.g. \citet{terekhov02}) as well as
stellar flares (e.g. \citet{mathio}).
Our preliminary analysis shows that quasi-periodic
oscillations seen in our numerical simulations
bear many similar features as the observed ones.
The key point is that the traditional explanation of these
oscillations in the observations involves {\it MHD waves}.
In the numerical simulations presented here, however, they 
are likely to be produced by standing sound waves 
caused by impulsive and localized
heating. Therefore, our explanation of these oscillations 
is purely hydrodynamic -- they are related to the
standing slow mode acoustic waves, similar to the observed 
by SUMER \citep{wang02}.
A detailed study of these quasi-periodic
oscillations will be presented elsewhere.

In a next step we plan to use the
outputs of this parametric study of hydrodynamic simulations,
which cover a wide parameter range of heating rates and 
heating time scales, as input for forward-fitting to the
observed physical parameters (densities and temperatures)
of the multi-loop flare on Bastille-day 2000 July 14.

\begin{acknowledgements}

This work was supported, in part,  by PPARC, UK.
Numerical calculations  were
performed using the PPARC funded MHD Cluster at St Andrews.
The work of MJA was partially supported by NASA contract
NAS5-38099 (TRACE).
\end{acknowledgements}


\begin{onecolumn}
\newpage
\begin{table}
\centering
\caption{Flare loop temperatures and densities as function of 
the heating rate
and heating duration, for a fixed loop  length of $L=55$ Mm, 
obtained from 1D hydrodynamic simulations.
Units of $E_{H0}$ are in (erg cm$^{-3}$ s$^{-1}$).
$\sigma_t$ is measured in seconds, while
temperatures and densities are given in
(MK) and  (cm$^{-3}$) respectively.}\label{tab1}
\begin{tabular}{llllll} \hline
Heating&
Heating&
Heating&
Temperature&
Temperature&
Electron\\
location&
rate&
duration&
at apex&
average&
density\\
Apex/Foot&
$E_{H0}$ &
$\sigma_t$ &
$T^{max}$ &
$T_{avg}$ &
$n_e^{max}/10^{11}$  \\ \hline \hline
A&     0.60&   41&    13.75&     6.78&     0.07\\
A&     0.60&   82&    13.86&     7.26&     0.11\\
A&     0.60&  164&    13.78&     7.44&     0.15\\
A&     0.60&  329&    13.67&     7.57&     0.17\\
F&     1.50&   41&    10.84&     6.35&     0.08\\
F&     1.50&   82&    10.80&     6.93&     0.14\\
F&     1.50&  164&    10.53&     7.06&     0.19\\
F&     1.50&  329&    10.11&     7.10&     0.25\\ \hline
A&     3.00&   41&    22.12&    11.80&     0.19\\
A&     3.00&   82&    22.07&    12.46&     0.29\\
A&     3.00&  164&    21.94&    12.72&     0.38\\
A&     3.00&  329&    21.80&    12.94&     0.43\\
F&     7.50&   41&    18.04&    11.52&     0.23\\
F&     7.50&   82&    17.64&    12.22&     0.41\\
F&     7.50&  164&    17.09&    12.37&     0.54\\
F&     7.50&  329&    16.51&    12.42&     0.69\\ \hline
A&    15.00&   41&    35.38&    20.38&     0.53\\
A&    15.00&   82&    35.18&    21.34&     0.94\\
A&    15.00&  164&    34.90&    21.53&     1.07\\
A&    15.00&  329&    34.95&    21.78&     1.09\\
F&    37.50&   41&    29.50&    20.57&     0.70\\
F&    37.50&   82&    29.28&    21.48&     1.33\\
F&    37.50&  164&    27.56&    21.36&     1.57\\
F&    37.50&  329&    26.88&    21.31&     1.90\\ \hline
A&    30.00&   41&    43.26&    25.72&     0.87\\
A&    30.00&   82&    42.99&    26.80&     1.43\\
A&    30.00&  164&    42.59&    26.87&     1.57\\
A&    30.00&  329&    42.64&    27.10&     1.64\\
F&    75.00&   41&    36.30&    26.29&     1.21\\
F&    75.00&   82&    37.65&    26.73&     1.95\\
F&    75.00&  164&    34.26&    26.74&     2.45\\
F&    75.00&  329&    33.15&    26.46&     3.11\\ \hline
A&    60.00&   41&    52.85&    32.42&     1.51\\
A&    60.00&   82&    53.44&    33.05&     2.10\\
A&    60.00&  164&    52.27&    33.29&     2.37\\
A&    60.00&  329&    52.27&    33.49&     2.48\\
F&   150.00&   41&    44.54&    33.52&     2.13\\
F&   150.00&   82&    45.31&    34.18&     2.97\\
F&   150.00&  164&    41.87&    33.05&     3.50\\
F&   150.00&  329&    40.19&    32.86&     4.19\\ \hline
\hline
\end{tabular}
\end{table}
\newpage
\begin{table}
\centering
\caption{Flare loop temperatures and densities as function 
of the loop length (in Mm),
for a fixed heating duration of $\sigma_t=329$ s.
Units of $E_{H0}$ are in (erg cm$^{-3}$ s$^{-1}$).
$\sigma_t$ is measured in seconds, while
temperatures and densities are given in
(MK) and  (cm$^{-3}$) respectively.}\label{tab2}
\begin{tabular}{llllll} \hline
Heating&
Heating&
Temperature&
Temperature&
Electron&
Loop\\
location&
rate&
at apex&
average&
density&
length\\
Apex/Foot&
$E_{H0}$ &
$T^{max}$ &
$T_{avg}$ &
$n_e^{max}/10^{11}$ &
$L$ [Mm] \\ \hline \hline
A&    60.00&      35.19&    28.18&     3.77&     13.75\\
A&    60.00&      42.12&    31.67&     3.30&     27.50\\
A&    60.00&      47.72&    32.88&     2.89&     41.25\\
A&    60.00&      52.27&    33.49&     2.48&     55.00\\
A&    60.00&      55.46&    33.94&     2.30&     68.75\\
A&    60.00&      58.77&    34.47&     2.17&     82.50\\
A&    60.00&      61.23&    34.74&     1.88&     96.25\\
A&    60.00&      63.57&    34.98&     1.83&     110.00\\ \hline
F&   150.00&      38.18&    31.84&     5.54&     13.75\\
F&   150.00&      39.33&    32.58&     4.61&     27.50\\
F&   150.00&      40.35&    32.89&     4.89&     41.25\\
F&   150.00&      40.19&    32.86&     4.19&     55.00\\
F&   150.00&      40.64&    32.85&     4.02&     68.75\\
F&   150.00&      40.86&    32.92&     3.85&     82.50\\
F&   150.00&      42.61&    33.24&     3.70&     96.25\\
F&   150.00&      43.13&    33.23&     3.56&     110.00\\ \hline
\hline 
\end{tabular}
\end{table}
\end{onecolumn}
\end{document}